\documentclass[twocolumn,prl]{revtex4}% Physical Review Letter
\usepackage{graphicx}    % Include figure files
\usepackage{dcolumn}     % Align table columns on decimal point
\usepackage{bm}          % bold math
\begin{document}
\preprint{Doppler}

%--- title ---
\title{Possible anomalous Doppler shift effect in
superconductor Sr$_{2}$RuO$_{4}$ }
%--- author ---
\author{
Y. Tanaka$^{1}$, K. Kuroki$^2$,  Y. Tanuma$^{3}$
and S. Kashiwaya$^{4}$
}%
%
%--- address ---
\affiliation{
$^1$Department of Applied Physics,
Nagoya University, Nagoya, 464-8603, Japan \\
$^2$
Department of Applied Physics and
Chemistry, The University of Electro-Communications,
Chofu, Tokyo 182-8585, Japan \\
$^3$Institute of Physics,
Kanagawa University,
Rokkakubashi, Yokohama, 221-8686, Japan \\
$^4$
National Institute of Advanced Industrial Science
and Technology, Tsukuba, 305-8568, Japan
}
%
%--- date ---
\date{\today}% It is always \today, today,
                %  but any date may be explicitly specified
%-----------------------------------------------------------
%   Abstract
%-----------------------------------------------------------
\begin{abstract}
The effect of the Doppler shift is studied in a model for the $\alpha$-$\beta$
bands of Sr$_2$RuO$_4$ consisting of two hybridized 1D bands.
Assuming a superconducting gap with nodes in the diagonal
directions, we examine the oscillation of the surface density
of states and the thermal conductivity under a rotating magnetic field.
Upon varying the strength of the hybridization, the oscillation
in these quantities is found to exhibit 2D to 1D crossover.
In the crossover regime, which corresponds to the actual Sr$_2$RuO$_4$,
the thermal conductivity exhibits a two-fold-symmetry oscillation,
while the four-fold-symmetry component in the oscillation is barely
detectable.
\end{abstract}
%-----------------------------------------------------------
                                % PACS, the Physics and Astronomy
                                % Classification Scheme.
%\keywords{Suggested keywords}%Use showkeys class option if keyword
                                %display desired
\maketitle
%-----------------------------------------------------------
%\section{Introduction}
%----------------------------------------------------------
A ruthenate superconductor
Sr$_{2}$RuO$_{4}$\cite{Maeno} has attracted much attention
as a possible candidate for spin-triplet superconductor.
After a theoretical prediction that the pairing should occur in the
spin-triplet channel,\cite{Rice}
several experiments \cite{Ishida,Duffy,Luke} have in fact supported
this possibility.
Early predictions were that the most probable $d$-vector is of the form
%$\Delta_{\sigma \sigma'}
%(\bm{k}) = i
%[ (\bm{d(k)} \bm(\sigma))\sigma^{y}]_{\sigma,\sigma'}$
%with
$\bm{d(k)}=(k_{x} \pm ik_{y})\hat{z}$.\cite{Rice}.
In this case, the gap does not have nodes, so the
material should be a gapful superconductor.
However, recent experiments
\cite{Nishizaki,Bonalde,Tanatar,Izawa1,Lupien}
indicate the existence of
nodes (or node-like structures) in the energy gap.
Stimulated by these experiments,
several theoretical models with gaps
having line nodes or node-like structures
have been proposed \cite{hasegawa,Zhitomir,won,Nomura,Kuroki,Miyake}.
On the other hand, thermal conductivity measurements under rotating
magnetic field\cite{Tanatar,Izawa1} have
indicated that the gap is almost isotropic within the planes,
suggesting indirectly the existence of {\it horizontal}
nodes in the gap, thereby excluding the possibility of
vertical nodes or node-like
structures proposed in some theoretical
studies.\cite{won,Nomura,Kuroki,Miyake}
These measurements under magnetic field have motivated our study.
%The point in this experiment is that the four-fold symmetry
%component in the oscillation of the thermal conductivity upon
%rotating the magnetic field is barely detectable in contradiction with
%model calculations which assumes gaps having four-fold symmetry.
%
%However, from microscopic theoretical viewpoints,
%since the electronic structure of Sr$_{2}$RuO$_{4}$ is strongly
%two-dimensional, it is unlikely to
%consider interlayer pairing which induce
%horizontal node.
%
\par
In the presence of a magnetic field,
it is known that
the energy spectrum of the quasiparticle is influenced
by the Doppler shift.\cite{FRS97,Vek99,Maki,Tewordt,Barash}
The Doppler shifted states around the nodes of the superconducting
gap contributes to the density of states at the Fermi energy.
%
%Due to this effect, thermal conductivity shows
%a characteristic oscillation as the direction of
%the magnetic field is rotated within the 2D plane, reflecting
%the nodal structure of the pair potential.
%
If we assume a free-electron-like Fermi surface, the density of states
is minimized (maximized) when the magnetic field is applied parallel to the
nodal (antinodal) direction, thereby exhibiting a four-fold symmetry
oscillation upon rotating the magnetic field.
This effect can be used to probe the
direction of the nodes by thermal conductivity
measurement,\cite{Izawa1,Maki,Tewordt,Yu,Aubin,Izawa2}
but since such experiments can be strongly affected by phonons,
we have recently proposed an alternative method :
magnetotunneling spectroscopy.\cite{Tanuma}
Namely, by rotating the magnetic field in the $ab$ plane,
the surface density of states (SDOS),
and thus the tunneling spectra, oscillates,
which enables us to determine the position of the nodes in the gap
without using the phase sensitive spectroscopy
based on the appearance of the Andreev bound states\cite{Kasi00}.
As a case study, we have considered the case of the high $T_c$ cuprates and
an organic superconductor $\kappa$-(BEDT-TTF)$_2$X, where we found that the
SDOS takes its minimum when the applied magnetic field is parallel to the
nodal direction, as in systems having free-electron-like Fermi surface.
It is not at all clear, however, whether this tendency holds
{\it regardless of the shape of the Fermi surface}.
\par
This is exactly where the present study sets in.
Here we study
the effect of the Doppler shift in the $\alpha$-$\beta$ bands,
the quasi one-dimensional(1D) bands, of Sr$_{2}$RuO$_{4}$.
This is motivated by some microscopic theories\cite{Kuroki,Nomura} which
proposes the presence of superconducting gaps in these bands
having nodes or node-like structures in the diagonal direction.
%
%We concentrate on these bands,
%which
Here we concentrate on the $\alpha$-$\beta$ bands, which 
implicitly assumes that a large nodeless gap
opens in the $\gamma$ band and
a small one in the $\alpha$-$\beta$ bands,
so that the main contribution to the
density of states at the Fermi level comes from the latter bands.
Considering a 2D model in which two 1D bands are hybridized,
and assuming a gap that has nodes
in the four diagonal directions (which of course corresponds to
{\it vertical} nodes in 3D systems),
we calculate the SDOS and the thermal conductivity
upon rotating the direction of the magnetic field.
To our surprise, we find that the four-fold-symmetry component in the
oscillation of these quantities, which should be clearly visible
in 2D systems having free-electron-like Fermi surface, can be barely
seen when the hybridization is moderate as in
the $\alpha$-$\beta$ bands of Sr$_{2}$RuO$_{4}$. What is even more striking
in this case is that
the thermal conductivity exhibits a strong two-fold-symmetry oscillation
reflecting the quasi-one dimensional nature of the
Fermi surface.
%
%When the magnitude of diagonal transfer integral $t'$,
%which induces the hybridization of the two one-dimensional
%bands, are sufficiently small,
%SDOS has a maximum (minimum) when the applied magnetic
%field is parallel to the nodal (anti-nodal) directions.
%
%While, with the increase of $t'$,
%SDOS and thermal conductivity
%has a minimum (maximum) when the applied magnetic
%field is parallel to the nodal (anti-nodal) directions,
%which are  consistent with  the conventional case.
%
%
%If we choose the magnitude of the hybridization
%as a plausible value to reproduce the Fermi surface of
%actual Sr$_{2}$RuO$_{4}$, the difference between the maximum
%and minimum value becomes very small and
%SDOS is quite insensitive to the direction of the
%magnetic field.
%
%The resulting thermal conductivity has a weak four fold symmetry
%by rotating the  the magnetic field.
%Then SDOS has a maximum (minimum) when the applied magnetic
%field is parallel to the nodal (anti-nodal) directions
%contradictory to the usual wisdom.
%
%In the light of our theory, thermal conductivity experiments
%does not exclude the vertical node model of pair potential
%in Sr$_{2}$RuO$_{4}$. \par
%
\par
%
%=======================================================
%\section{Formulation}
%=======================================================
%
%A schematic illustration of our
%quasi-2D superconductors with two quasi-one dimensional
%bands and the direction of the magnetic field is shown in
%Fig. 1 where the magnetic field is
%rotated within the $xy$-plane.

%----- figure 1 ------
\begin{figure}[htb]
\begin{center}
\scalebox{0.8}{
\includegraphics[width=7cm,clip]{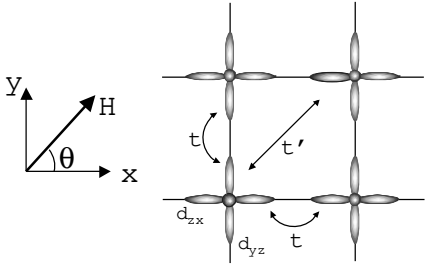}}
\caption{
Left: magnetic field $H$ in the $\theta$ direction.
Right: the 2D model considered in the present study.
\label{fig:01}
}
\end{center}
\end{figure}
%---------------------
%
First let us focus on the tunneling spectrum, namely the
SDOS. Assuming that the penetration depth is much
longer than the coherence length,
the vector potential can be expressed as
$\bm{A}(\bm{r})=(H\lambda e^{z/\lambda}\sin \theta,-H\lambda 
e^{z/\lambda}\cos \theta,0)$,
where  $\theta$ is the angle of
the magnetic field measured from the $x$-axis as shown in Fig.\ref{fig:01}
and $\lambda$ is the penetration length, respectively.
In the following, we assume $\lambda$ is much larger than
the coherence length $\xi$ and
$\bm{A}(\bm{r})$ can be approximated as
$\bm{A}(\bm{r}) \sim \bm{A}_{0}
=(H\lambda \sin \theta,-H\lambda \cos \theta,0)$.
Then the quasiparticle momenta $k_x$ and $k_y$
in the $x$ and $y$ directions can be given
as $\tilde{k}_{x} = k_x +(H/\pi \xi H_{0})\sin \theta$
and $\tilde{k}_{y} = k_y -(H/\pi \xi H_{0})\cos \theta$,
where $H_{0}=\phi_{0}/(\pi^{2} \xi \lambda)$
with $\phi_{0}=h/(2e)$.
Thus, the SDOS at zero energy (the Fermi energy)
can be expressed as, \cite{Kasi00}
\begin{eqnarray}
  \rho(\theta,H) &=&
  \int ^{\infty}_{-\infty}d\omega
\bar{\rho}_{\rm S}(\omega)
{\rm sech}^{2}\left ( \frac{\omega}{2k_{\rm B}T} \right ),
\\
\bar{\rho}_{\rm S}(\omega) &=& \frac{1}{2}
\sum_{\bm{k},\sigma} \left \{ |u_{\bm{k},\sigma}|^{2}
\left [
  \delta(\omega - E_{\bm{k},\sigma})
+\delta(\omega - E_{-\bm{k},\sigma}) \right ]
\right .
\nonumber \\
&& \left .
+|v_{\bm{k},\sigma}|^{2} \left [
  \delta(\omega + E_{ \bm{k},\sigma})
+\delta(\omega + E_{-\bm{k},\sigma})
\right ]
\right \}.
\end{eqnarray}

\par

\begin{eqnarray}
E_{\pm \bm{k},\sigma}=\frac{ \left [
(\xi_{\bm{k,\sigma}}-\xi_{-\bm{k},\sigma})
\pm \sqrt{(\xi_{\bm{k},\sigma}+\xi_{-\bm{k},\sigma})^{2}
+4|\Delta_{\bm{k},\sigma}|^{2}}
\right ]}{2},
\end{eqnarray}
\begin{eqnarray}
|u_{\bm{k},\sigma}|^{2}=\frac{1}{2}
\left (1+\frac{\eta_{\bm{k},\sigma}}{\Gamma_{\bm{k},\sigma}} \right ),
\quad
|v_{\bm{k},\sigma}|^{2}=\frac{1}{2}
\left (1-\frac{\eta_{\bm{k},\sigma}}{\Gamma_{\bm{k},\sigma}} \right ),
\end{eqnarray}
with $\eta_{\bm{k},\sigma} = \xi_{\bm{k},\sigma}+\xi_{-\bm{k,\sigma}}$
and $\Gamma_{\bm{k},\sigma} = \sqrt{\eta_{\bm{k},\sigma}^{2}
+4|\Delta_{\bm{k},\sigma}|^{2}}$,
with band index $\sigma(=\alpha$  or $\beta)$.
%
%---------------------------------------
The two bands $\alpha$ and $\beta$ result from the
hybridization of two 1D bands, where the energy dispersion
$\xi_{\bm{k},\alpha}$ and $\xi_{\bm{k},\beta}$ are given as follows.
\begin{equation}
\xi_{\bm{k},\alpha}=
\frac{1}{2}
\{ (\epsilon_{kxz} + \epsilon_{kyz})
- \sqrt{ (\epsilon_{kxz} - \epsilon_{kyz})^{2} + 4g_{\bm{k}}^{2} }
\}
\end{equation}

\[
\xi_{\bm{k},\beta}=
\frac{1}{2}
\{ (\epsilon_{kxz} + \epsilon_{kyz})
+ \sqrt{ (\epsilon_{kxz} - \epsilon_{kyz})^{2} + 4g_{\bm{k}}^{2} }
\}
\]
\[
\epsilon_{kxz}= -2t \cos(\tilde{k}_{x}a) -\mu, \
\epsilon_{kyz}= -2t \cos(\tilde{k}_{y}a) -\mu, \
\]
\[
g_{\bm{k} }=-4t'\sin(\tilde{k}_{x}a)\sin(\tilde{k}_{y}a)
\]
where $\epsilon_{kxz}$ and $\epsilon_{kyz}$ are
the dispersion of the $d_{xz}$ and $d_{yz}$ bands, respectively
(see Fig.\ref{fig:01}).
$t$ and $\mu$ are fixed at $t=0.18t_{0}$ and $\mu=0.17t_{0}$,
where $t_0$ is the unit of the energy about 1eV,
while $t'$ is varied as a key parameter in the present study
which controls the
strength of the hybridization between the two bands.
Appropriate value of $t'$ for Sr$_2$RuO$_4$
should be $t'=0.01\sim 0.02t_{0}$,
but $t'$ is varied in a wider range to
see the crossover between 1D and 2D.
\par
As for the gap functions,
its absolute value (note that only the absolute value enters
in eqns.(3)\&(4))
are chosen as
\begin{equation}
|\Delta_{\bm{k},\alpha}|=|\Delta_{\bm{k},\beta}|
= \Delta_{0}|(\cos(k_x)-\cos(k_y))|,
%[ \frac{\exp[-\alpha( \mid k_{x} \mid -Q)^{2}] }{\Sigma_{\alpha,ky}}
%+ \frac{\exp[-\alpha( \mid k_{y} \mid -Q)^{2}] }{\Sigma_{\alpha,kx}}]
\end{equation}
in order to take
  into account the node-like structures
of the spin-triplet gap
functions (having $p_x+ip_y$ symmetry) found in refs.\cite{Kuroki,Nomura}.
Although we assume this phenomelogical form for simplicity,
our conclusion is 
not qualitatively affected by the detailed form of the gap,
that is, only the direction of the nodes is important.
%\begin{equation}
%\Delta_{\bm{k},\beta}
%= \Delta_{0}
%[ \frac{\exp[-\alpha( \mid k_{x} \mid -Q)^{2}] }{\Sigma_{\beta,ky}}
%+ \frac{\exp[-\alpha( \mid k_{y} \mid -Q)^{2}] }{\Sigma_{\beta,kx}}]
%\end{equation}
%
%\[
%\Sigma_{\alpha,kx(ky)}=1 + \exp[-\alpha(\mid k_{x}(k_{y}) \mid -\beta Q)], \
%\]
%\begin{equation}
%\Sigma_{\beta,pkx(ky)}=1 + \exp[-\alpha(\mid k_{x}(k_{y}) \mid -\gamma Q)], \
%\end{equation}
%with $\alpha=7$, $\beta=0.8$, $\gamma=1.2$
%and $Q=\frac{2\pi}{3}$, respectively.
%
%This pair potential has four vertical node like structures
%around $(\pm 2\pi/3,  \pm 2\pi/3)$. \par
%
In the following calculation,
we fix the parameters $\Delta_{0}=0.05t_{0}$ and $T=0.002t_{0}$,
but our conclusion is not qualitatively affected by the choice of
these values.
\par
\begin{figure}[htb]
\begin{center}
\scalebox{0.45}{
\includegraphics[width=10cm,clip]{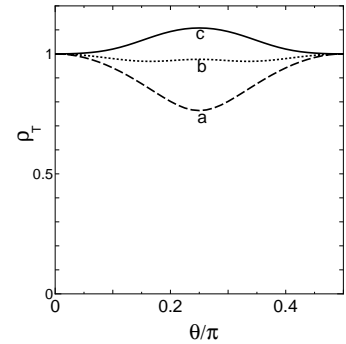}}
\caption{
The normalized SDOS plotted as a function of $\theta$
for $H=0.15H_{0}$ with
a: $t^{\prime}/t=0.1$,
b: $t'/t=0.01$, and c: $t'/t=0.005$.
\label{fig:02}}
\end{center}
\end{figure}

In Fig.\ref{fig:02}, we plot the normalized SDOS
$\rho_{T}=\rho(\theta,H)/\rho(0,H=0.15H_{0})$ as a function of $\theta$
for several values of $t'$.
In the case of $t'=0.1t_{0}$, where the hybridization is strong and
the Fermi surface is round,
$\rho_{T}$ is minimized at $\theta=\pi/4$, namely
the nodal direction,
and maximized at $\theta=0,\pi/2$, the antinodal
direction,
which is consistent with the previous theories for 2D systems.\cite{Vek99}
By contrast, when the hybridization is weak and the bands are
essentially 1D ($t'=0.005t_{0}$),
the maximum and the minimum of the oscilation is entirely reversed.
Consequently, for a moderate hybridization $t'=0.01t_{0}$,
the SDOS becomes almost constant upon rotating the magnetic field.
\par
%
%----- figure 2 ------
\begin{figure}[htb]
\begin{center}
\scalebox{0.8}{
\includegraphics[width=7cm,clip]{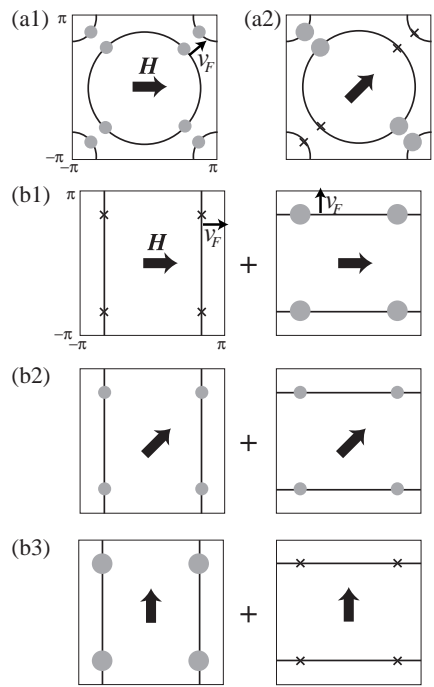}}
\caption{
The states (hatched circles) around the nodes of the gap
contributing to the density of states when
$t'$ is sufficiently large and
the magnetic field is applied in
(a1) the antinodal or (a2) the nodal directions, or
%In this case, contribution to the
%SDOS becomes the  summation from
%$\epsilon_{xz}$ band and $\epsilon_{yz}$
%band.
when $t'=0$ and
the magnetic field is applied in the
(b1)\& (b3) antinodal or (b2) the nodal directions.
%when the hybridization is sufficiently large.
%In this case, $\alpha$ and $\beta$ band form
%quasi-two dimensional Fermi surface.
The size of the circles schematically represent the
magnitude of the contribution.
\label{fig:03}
}
\end{center}
\end{figure}

%---------------------
%
This crossover between large and small $t'$ can be understood as
follows.
Let us first note that the Doppler shift is essentially given by
$ \bm{v}_{F} \cdot \bm{A}_{0}$,
where $\bm{v}_F$ is the Fermi velocity \cite{FRS97}
and the vector potential $\bm{A}_{0}$ is perpendicular
to the magnetic field $\bm{H}$.
%In Fig. \ref{fig:1a}, we show the schematic illustration of the region
%contributing to SDOS for the antinodal and nodal orientation of the
%magnetic field.
%
When the hybridization is sufficiently strong,
the Fermi surface is round as in
Figs. \ref{fig:03}(a1) and \ref{fig:03}(a2), so that the situation
is the same as
in the previous studies for 2D systems.\cite{Vek99}
Namely, all the states around the diagonal nodes
have non-zero $v_{F,x}$ and $v_{F,y}$,
thereby  contributing to the density of states
when the magnetic field is applied in the antinodal direction
(Fig. \ref{fig:03}(a1)).
When the field is applied in the nodal direction, on the other hand,
the states around the nodes parallel to the field
do not contribute to the density of states,
while the contribution of the states around the nodes
perpendicular to the field
is only a factor of $\sqrt{2}$ larger than it is when the
field is applied in the antinodal direction
(Fig.\ref{fig:03}(a2)).
Thus, the SDOS  becomes larger when the field is applied in the
antinodal direction.
\par
The situation completely changes when we
consider the weak hybridization limit, $t'=0$, where
the two one-dimensional bands contribute independently to
the density of states.
In this case,
when the magnetic field is applied along the antinodal direction
(Fig.\ref{fig:03}(b1)\&(b3)), only the states on the Fermi surface parallel to
the field gives contribution.
On the other hand, for a field in the nodal direction,
states on both Fermi surfaces have contribution that is
a factor of $\sqrt{2}$ smaller than it is when the
field is in the antinodal direction (Fig.\ref{fig:03} (b2)).
Consequently, in the 1D limit, the SDOS becomes larger
when the field is applied in the {\it nodal} direction.
%
%----- figure 3 ------
%---------------------
%
\par
\begin{figure}[htb]
\begin{center}
\scalebox{0.7}{
\includegraphics[width=10cm,clip]{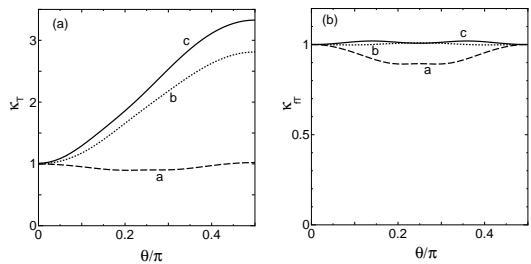}}
\caption{
(a)$\kappa_{T}$ and (b)$\kappa_{fT}$ plotted as functions of $\theta$ for
$H=0.15H_{0}$.
a: $t^{\prime}/t=0.1$,
b: $t'/t=0.01$, and c: $t'/t=0.005$.
\label{fig:04}}
\end{center}
\end{figure}

%
%---------------------
%
We now move on to the thermal conductivity.
Thermal conductivity $\kappa_{xx}(\theta,H)$ can be
expressed as \cite{Agterberg,Arfi}
\begin{equation}
\kappa_{xx}(\theta,H)=-\sum_{{\bm k},\sigma}
{\rm sech}^{2}\left ( \frac{E_{\bm{k},\sigma} }{k_{B}T} \right )
E_{\bm{k},\sigma}^{2}
v_{\bm{k},x}^{2} \tau_{\bm{k},\sigma}
\label{thermalcond}
\end{equation}
with $v_{\bm{k},x}=(\partial E_{\bm{k},\sigma}/\partial k_{x})$,
where we assume a constant $\tau_{\bm{k},\sigma}$
with $\tau_{\bm{k},\sigma}=\tau_{0}$.
As for the energy dispersion $E_{\bm k}$, we assume the form
adopted for the calculation of SDOS for simplicity.
Although this may not accurately correspond to
the actual experimental situation, we believe that
the essential physics can be captured within this
formalism.
The thermal conductivity for $H=0.15H_{0}$,
normalized as $\kappa_{T}=\kappa_{xx}(\theta,H)/\kappa_{xx}(0,H)$,
is shown in Fig. \ref{fig:04}(a).
For $t'=0.1t_{0}$, $\kappa_{T}$ is minimized around
$\theta=\pi/4$ and the oscillation essentially has a four-fold symmetry,
which is consistent with the previous studies for 2D systems.\cite{won}

By contrast, for $t'=0.005t_{0}$ and $t'=0.01t_{0}$,
$\kappa_{T}$ exhibits a strong two-fold-symmetry oscillation,
taking its  maximum at $\theta=\pi /2$ and a minimum at $\theta=0$.
The anomalous two-fold-symmetry oscillation
in the case of weak to moderated hybridization 
can be understood by considering again the
$t'=0$ limit.
Namely, when $\theta=0$,
only the states on the $d_{yz}$ branch of the Fermi surface,
where $v_{{\bm k}x}=0$, gives contribution to the
density of states (see Fig.\ref{fig:03}(b1)),
so that $\kappa_{xx}(\theta,H)\simeq 0$ according to eq.(\ref{thermalcond})
at zero temperature.
When $\theta=\pi/2$, on the other hand, the states on the
$d_{xz}$ branch contribute to the density of states
(Fig.\ref{fig:03}(b3)),
so that $\kappa_{xx}(\theta,H)$ becomes large.
%As seen from right the panel of Fig. 3(a1), the resulting magnitude of
%$v_x$  is almost zero (exact zero for $t'=0$) and
%contirubution from right panel is negligible.
%
%On the other hand, since $H$ is paralell to the direction of Fermi velocity
%around node, the contribution from left panel is negligible.
%
%However, upon rotating $H$ up to $\pi/2$, the contribution from left panel
%becomes significant since
%direction of $v_{x}$ is perpendicular to $H$ and $v_{x}$ has appreciable
%magnitude.
%
%This is the origin of the two fold symmetry of $\kappa_{T}$
%which is specific to quasi one dimensional electronic structures.
%\par
%\par
%
%----- figure 4 ------

Let us now look into the four-fold-symmetry
component of  $\kappa_{xx}(\theta,H)$.
Since the two-fold-symmetry
components in $\kappa_{xx}(\theta,H)$ and $\kappa_{yy}(\theta,H)$
have the same absolute values and the opposite signs to each other,
we can focus on the four-fold-symmetry component
by looking into $\kappa_{fT}$ defined as
$\kappa_{fT}=\kappa_{f}(\theta,H)/\kappa_{f}(0,H)$
\[
\kappa_{f}(\theta,H)
=  \frac{1}{2}[\kappa_{xx}(\theta,H) + \kappa_{yy}(\theta,H)].
\]
As seen in Fig.\ref{fig:04}(b), the $\theta$ dependence of $\kappa_{fT}$
qualitatively resembles that of the SDOS shown in Fig.\ref{fig:02}.
Here again, the four-fold-symmetry component
in the oscillation of $\kappa_{xx}(\theta,H)$ can be barely seen
for moderate (and realistic) values of $t'$.
The absence of (or very small) four-fold-symmetry component
is consistent with the experimental observations
in refs.\cite{Tanatar,Izawa1}, but
the angles where $\kappa_{xx}(\theta,H)$ is maximized and minimized
are reversed compared to the two-fold-symmetry
oscillation found in ref.\cite{Izawa1}.
However, since we have studied the case of $T\ll\Delta_{0}$
considering that the effect of phonons can be neglected
at low temperatures in the actual experiments,
we believe that further experiments at temperatures
much lower than the energy scale of the superconducting gap is necessary to
verify our prediction for vertical diagonal nodes.
\par
%
%where $\kappa_{xx}(\theta,H)$ and $\kappa_{yy}(\theta,H)$
%can be expressed as
%$\kappa_{xx}(\theta,H)=\sum_{n} a_{n}\cos(2n\theta)$ and
%$\kappa_{yy}(\theta,H)=\kappa_{xx}(\pi/2-\theta,H)$.
%Then $\kappa_{f}(\theta,H)$ can be expressed as
%\[
%\kappa_{f}(\theta)=\sum_{n} a_{n}\cos(4n\theta)
%\]
%----- figure 4 ------
%\begin{figure}[htb]
%\begin{center}
%\scalebox{0.4}{
%\includegraphics[width=10cm,clip]{fig5.eps}}
%\caption{
%$\kappa_{fT}$ is plotted as a function of $\theta$ for
%$H=0.15H_{0}$.
%a $t^{\prime}/t=0.005$,
%b $t'/t=0.01$, and c $t'/t=0.1$.
%\label{fig:05}}
%\end{center}
%\end{figure}
%---------------------
%==================================================
%\section{Conclusions}
%==================================================
%
To summarize, we have studied the effect of the Doppler shift
on SDOS and the thermal conductivity
in a 2D model consisting of hybridized 1D bands,
assuming a superconducting gap that has nodes in the diagonal directions.
When the hybridization is strong and the Fermi surface is round,
both the SDOS and $\kappa_{xx}(\theta,H)$ exhibits a four-fold-symmetry
oscillation, taking a  maximum (minimum)
when the magnetic field is applied parallel to the
antinodal (nodal) directions, which is consistent with previous
theories for 2D systems.
By contrast, when the hybridization is very weak,
the SDOS 
is maximized (minimized) when the magnetic
field is applied parallel to the nodal (anti-nodal) directions.
What is more remarkable in the case of weak to moderate
hybridization is that
the thermal conductivity exhibits a strong two-fold-symmetry
oscillation,
reflecting the quasi-1D nature of the Fermi surface.
Since the actual Sr$_2$RuO$_4$ corresponds to the regime with
moderate hybridization,
$\kappa_{xx}(\theta,H)$ should exhibit a two-fold-symmetry
oscillation, taking its maximum (minimum) when the magnetic field is applied
in the $y$ $(x)$ direction,
while the four-fold-symmetry component in
the SDOS and $\kappa_{xx}(\theta,H)$ should be barely detectable.
%In the light of the present theory,
%the absence of clear four fold symmetry
%in the thermal conductivity experiments \cite{Izawa1}
%does not necessarily
%support the presence of the horizontal node model.
\par
%%
%==================================================
Y.T. would like to thank
Y. Matsuda for useful and fruitful discussions.
%
%This work was supported by
%the Core Research for
%Evolutional Science and Technology (CREST)
%of the Japan Science
%and Technology Corporation (JST).
%
%The computational aspect of this work has been performed
%at the facilities of the Supercomputer Center,
%Institute for Solid State Physics,
%University of Tokyo and the Computer Center.
%======Reference===================================
%

%===============================================================

\begin{thebibliography}{99}
%
\bibitem{Maeno} Y.~Maeno, $et$ $al.$
%H.~Hashimoto, K.~Yoshida, S.~Nishizaki, T.~Fujita,
% J.~G.~Bednorz, and F.~Lichtenberg,
Nature \textbf{372}, 532 (1994).
%
\bibitem{Rice} T.~M.~Rice and M.~Sigrist, J. Phys. Condens. Matter 
\textbf{7}, L643 (1995).
%
\bibitem{Ishida}  K.~Ishida, $et$ $al.$
%H.~Mukuda, Y.~KItaoka, K.~Asayama, Z.~Q.~Mao, Y.~Mori and Y.~Maeno,
Nature \textbf{396}, 658 (1998); Phys. Rev. Lett. {\bf 84}
5387 (2000).
%
\bibitem{Duffy} J.A. Duffy $et$ $al.$, Phys. Rev. Lett. {\bf 84}
5387 (2000).
\bibitem{Luke} G. M. Luke $et$ $al.$ Nature {\bf 394} 558 (1998).
%
%%%%%%%%%%%%%%%%%%%%%%%%%%%%%%%%%%%%%%%%%%%%%%
% Interesting phenomena (tunneling et al)
%%%%%%%%%%%%%%%%%%%%%%%%%%%%%%%%%%%%%%%%%%%%%
%\bibitem{yamashiro} M.~Yamashiro, Y.~Tanaka and S.~Kashiwaya: Phys. Rev. B
%\textbf{56} (1996) 7847; Yamsiro $et$ $al$,
%J. Phys. Soc. Jpn. \textbf{67}, 3224 (1998).
%\bibitem{Laube}
%F. Laube, $et$ $al.$
%G. Goll, H.v. L\"{o}hneysen,
%M. Fogelstr\"{o}m, and F. Lichtenberg:
%Phys. Rev. Lett. {\bf 84} [2000] 1595.
%
%\bibitem{Mao}
%Z.Q. Mao, $et$ $al.$
%K.D. Nelson, R. Jin, Y. Liu,
%and Y. Maeno:
%Phys. Rev. Lett. {\bf 87} [2001] 037003.
%
%%%%%%%%%%%%%%%%%%%%%%%%%%%%%%%%%%%%%%%%%%%%%%%
%  Experiments showing the exisetence of node
%%%%%%%%%%%%%%%%%%%%%%%%%%%%%%%%%%%%%%%%%%%%%%
\bibitem{Nishizaki} S. Nishizaki, Y. Maeno and Z. Q. Mao,
J. Phys. Soc. Jpn. {\bf 69} 572 (2000).
\bibitem{Bonalde}
I. Bonalde $et$ $al.$ Phys. Rev. Lett. {\bf 85} 4775 (2000).
%
\bibitem{Tanatar}
M. A. Tanatar $et$ $al.$, Phys. Rev. B {\bf 63} 064505 (2001);
Phys. Rev. Lett. {\bf 86} 2649 (2001).
%
\bibitem{Izawa1}
K. Izawa $et$ $al.$ Phys. Rev. Lett. {\bf 86} 2653 (2001).
%
\bibitem{Lupien}
C. Lupien $et$ $al.$, Phys. Rev. Lett. {\bf 86} 5986 (2001).
%
%%%%%%%%%%%%%%%%%%%%%%%%%%%%%%%%%%%%%%%%%%%%%%%%%%5
%  Horizontal Node
%%%%%%%%%%%%%%%%%%%%%%%%%%%%%%%%%%%%%%%%%%%%%%%%%%
\bibitem{hasegawa} Y.~Hasegawa, K.~Machida and M.~Ozaki,
  J. Phys. Soc. Jpn. \textbf{69}, 336 (2000).
%
\bibitem{Zhitomir}
M.E. Zhitomirsky and T.M. Rice, Phys. Rev. Lett. {\bf 87} 057001 (2001).
%
%%%%%%%%%%%%%%%%%%%%%%%%%%%%%%%%%%%%%%%%%%%%%%%%%%%
% Verical node (node like)
%%%%%%%%%%%%%%%%%%%%%%%%%%%%%%%%%%%%%%%%%%%%%%%%%%%%%
\bibitem{won}
H.~Won and K.~Maki, Europhys. Rev. Lett.\textbf{52}, 427 (2000).
T. Dahm, H. Won and K. Maki, cond-mat/0006301.
%
\bibitem{Nomura}
T. Nomura and K. Yamada, J. Phys. Soc. Jpn. \textbf{69}, 3678 (2002);
J. Phys. Soc. Jpn. \textbf{69}, 404 (2002).
%
\bibitem{Kuroki}
K. Kuroki, M. Ogata, R. Arita, and H. Aoki,
Phys. Rev. B \textbf{63}, 060506 (2002).
%
\bibitem{Miyake}
K.~Miyake and O.~Narikiyo, Phys. Rev. Lett.\textbf{83}, 1423 (1999).
%
%%%%%%%%%%%%%%%%%%%%%%%%%%%%%%%%
% Dppler shift
%%%%%%%%%%%%%%%%%%%%%%%%%%%%%%%%%
%
\bibitem{FRS97}
M. Fogelstr\"{o}m, D. Rainer, and J. A. Sauls,
Phys. Rev. Lett. {\bf 79}  281 (1997).
%
\bibitem{Vek99}
I. Vekhter,$et$ $al.$
%P.J. Hirschfeld,
%J.P. Carbotte, and E.J. Nicol,
Phys. Rev. B {\bf 59}  R9023 (1999).

%
\bibitem{Maki}
K. Maki, G.L. Yang, and H. Won,
Physica C {\bf 341-348}  1647 (2000),
H. Won and K. Maki,
cond-mat/0004105 (unpublished).
%
\bibitem{Tewordt}
L. Tewordt and D. Fay, Phys. Rev. B {\bf 64}
024528 (2001).
%
\bibitem{Barash}
Yu. S. Barash, A. A. Svidzinskii and V. P. Mineev, 
Pis'ma Zh. Eksp. Teor. Fiz. {\bf 65} 606 (1997) 
[JETP Lett. {\bf 65} 25 (1997)]. 
%
\bibitem{Yu}
F. Yu, $et$ $al.$
%M.B. Salamon, A.J. Leggett, W.C. Lee,
%and D.M. Ginsberg,
Phys. Rev. Lett. {\bf 74} 5136 (1995),
{\bf 75}  3028 (1995).
%
\bibitem{Aubin}
H. Aubin, $et$ $al.$
%K. Behnia, M. Ribault, R. Gagnon,
%and L. Taillefer,
Phys. Rev. Lett. {\bf 78}  2624 (1997).
%
\bibitem{Izawa2}
K. Izawa $et$ $al.$,
Phys. Rev. Lett. {\bf 87}  057002 (2001);
Phys. Rev. Lett. {\bf 88}  027002 (2002).
%
%%%%%%%%%%%%%%%%%%%%%%%%%%%%%%%%%%%%%%%%%%%%%%%%%%%%%
% Magneto tunneling
%%%%%%%%%%%%%%%%%%%%%%%%%%%%%%%%%%%%%%%%%%%%%%%%%%%%%
\bibitem{Tanuma}
Y. Tanuma, $et$ $al$,
Phys. Rev. B {\bf 60} 094507 (2002);
Phys. Rev. B {\bf 60} 174502 (2002);
Y. Tanaka $et$ $al.$ J. Phys. Soc. Jpn. {\bf 66}
2102 (2002).
%%%%%%%%%%%%%%%%%%%%%%%%%%%%%%%%%%%%%%%
% Andreev bound statea
%%%%%%%%%%%%%%%%%%%%%%%%%%%%%%%%%%%%%%%%%
\bibitem{Kasi00}
Y. Tanaka and S. Kashiwaya,
Phys. Rev. Lett. {\bf 74}, 3451 (1995);
S. Kashiwaya and Y. Tanaka,
Rep. Prog. Phys. {\bf 63}, 1641 (2000); 
T. L\"{o}fwander, V. S. Shumeiko, and G. Wendin, 
Supercond. Sci. Tech. {\bf 14} R53 (2001). 
%%%%%%%%%%%%%%%%%%%%%%%%%%%%%%%%%%%%%%%%
% Expression of thermal conductivity
%%%%%%%%%%%%%%%%%%%%%%%%%%%%%%%%%%%%%%%%
%
\bibitem{Agterberg}
D.F. Agterberg, T.M. Rice and M. Sigrist, Phys. Rev. Lett.
{\bf 78} 3374 (1997).
\bibitem{Arfi}
B. Arfi and C.J. Pethick,
Phys. Rev. B {\bf 38} 2312 (1988).
%%%%%%%%%%%%%%%%%%%%%%%%%%%%%%%%
% Dppler shift
%%%%%%%%%%%%%%%%%%%%%%%%%%%%%%%%%
%

%
%\bibitem{Izawa1}
%K. Izawa,
%H. Takahashi, H. Yamaguchi, Y. Matsuda,
%M. Suzuki, T. Sasaki, T. Fukase, Y. Yoshida, R. Settai,
%and Y. Onuki,
%Phys. Rev. Lett. {\bf 86} [2001] 2653.
%
%\bibitem{Izawa2}
%K. Izawa,
%H. Yamaguchi, Y. Matsuda, H. Shishido,
%R. Settai, and Y. Onuki,
%Phys. Rev. Lett. {\bf 87} [2001] 057002.
%
%\bibitem{Izawa3}
%K. Izawa,
%H. Yamaguchi, T. Sasaki, and Y. Matsuda,
%Phys. Rev. Lett. {\bf 88} [2002] 027002.
%
\end{thebibliography}
\end{document}